\documentclass[manuscript,screen]{acmart}
\AtBeginDocument{%
  }

\setcopyright{acmcopyright}
\copyrightyear{2022}
\acmYear{2022}
\acmDOI{XXXXXXX.XXXXXXX}

\acmConference[RECSYS '22]{The ACM Conference Series on Recommender Systems}{September 18--23,
  2018}{Seattle, WA}
\acmPrice{15.00}
\acmISBN{978-1-4503-XXXX-X/18/06}

\begin{document}
\title{Augmenting Netflix Search with In-Session Adapted Recommendations}

\author{Moumita Bhattacharya}
\email{mbhattacharya@netflix.com}
\orcid{0000-0002-7836-4504}
\author{Sudarshan Lamkhede}
\orcid{0000-0001-8699-3776}
\email{slamkhede@netflix.com}
\affiliation{%
  \institution{Netflix Research}
  \city{Los Gatos}
  \state{California}
  \country{USA}
}

\renewcommand{\shortauthors}{Bhattacharya and Lamkhede, et al.}

\begin{abstract}
We motivate the need for recommendation systems that can cater to the members’ in-the-moment intent by leveraging their interactions from the current session. We provide an overview of an end-to-end in-session adaptive recommendations system in the context of Netflix Search. We discuss the challenges and potential solutions when developing such a system at production scale. 
\end{abstract}



\maketitle

\section{Motivation}
Recommender systems are a critical component of many online products and services, as they help members discover the most relevant products that they are interested in from a large catalog. However, building such systems to continually produce relevant items is a challenging task because members’ tastes evolve over time and their multifaceted interests and needs may be different at different times. Surfacing the right product to the right member at the right time requires the recommender systems to constantly adapt to members’ shifting needs. 

In the context of streaming services such as Netflix, a member typically searches for a title when they have something specific they are looking for (\textit{fetch and find intent} \cite{lamkhede2019challenges}) or if they are exploring the catalog further (\textit{explore intent} \cite{lamkhede2019challenges}). It could be a delightful member experience if for such members we can identify these search intents (fetch, find or explore) and show relevant recommendations before they have to explicitly start typing queries. We call this \textit{pre-query recommendations}, though has also  been referred to as anticipatory search \cite{liebling2012anticipatory}. Consider a member who often watches reality TV shows but for one session she is interested in watching a Korean horror movie. Let’s say the member just arrived at Netflix search after interacting with a few Korean titles and horror titles elsewhere on the platform. At this stage, if we can predict her in-the-moment intent, we can surface relevant pre-query recommendations without her needing to provide any explicit input. That can be a delightful experience and help her find something she would like to watch now without much effort on her part.

On our product, many members engage with the homepage and other pages before navigating to Search. These search sessions \footnote{A session can be defined as a continuous period of member engagement with the platform bounded by inactivity for a certain period (e.g. 30 minutes).} have the opportunity to use the member’s recent browsing signals within that session to generate recommendations on the pre-query canvas. Unlike most recommender systems that have been accurately learning a member’s interest profile based on her past viewing habits and based on behavior of similar members \cite{steck2021deep, hidasi2015session}, this requires an approach that not only learns members’ long-term preferences but also utilizes their short term (\textit{in-the-moment}) preferences, yielding in-session adapted recommendations. Notably, a previous work by Wu et al. \cite{wu2016using} has shown that leveraging such within-session member navigation patterns for adapting recommendations in real-time generates more relevant rankings. Furthermore, in-session adapted recommendations can provide useful insights for members with insufficient historical viewing preferences.

Based on all the above, we hypothesize that approaches that can adapt to capture a member’s in-the-moment preference that are not only aware of a member’s long-term preference as before but are also aware of their short-term intent, will produce more relevant recommendations on the search pre-query canvas by anticipating what the member intent to \textit{search next}. We thus experiment with a large-scale ranking system that pays attention to what members are doing within the current session while still remembering the member's past preferences.  Although we leverage both in-session and cross-session signals, for brevity we will only refer to \textit{in-session} in the discussion. In this talk, we plan to provide an overview of the machine learning approaches and the infrastructure considerations to generate such in-session adapted recommendations for the pre-query canvas.

\section{Related Work}
Session-based recommendations have been an active area of research in recent years, where sequential models such as RNN \cite{hidasi2015session} and Attention architectures \cite{kang2018self} have been used to model variable length sequences that leverage member's actions on a platform. However, in these studies capturing members’ near-real-time intent and balancing with long-term interests was not the focus, as is in our case. In-session recommendation has also been approached with attention techniques that focus on specific parts of the member engagement sequence. For example, in the study by Liu et al. \cite{liu2018stamp} a short-term attention model was developed to jointly learn members’ general preference using historical sequence, while specifically paying attention to the most recent engagement from the current session. Additional approaches such as continual learning \cite{mi2020ader} and reinforcement Learning \cite{chen2021user}, have shown to be beneficial for such recommendations. 


\section{Approach}
To serve a truly real-time, in-session-adapted recommender we need a lot of aspects in the production system to come together such as \textit{just-in-time} (JIT) server call patterns, client side logging, server side logging, accessibility of members’ near real-time browsing signals, and a machine learning model that can quickly handle real-time context to make predictions about a member’s short-term intent. In this talk, we plan provide an overview of these different aspects. 

\underline{\textit{Data Infrastructure}}: One of the key requirements to design an in-session adaptive recommender system is to fetch the ranked results as close to when the member will see them, so that all the within-session engagements until then can be taken into account. To achieve this we need to have live requests to the server triggered when the member is about to see the recommendations. That is, when a member comes to the page, the client would need to make a call to the service at that moment and within acceptable latency to render fresh recommendations. Moreover, recommendations can not be retrieved from a previous cached result as we want fresh recommendations to be generated by using the actions that a member undertook on the platform thus far in the current session. Additionally, the member interactions in a given session need to be passed to the machine learning model in a near-real-time fashion such that the model inference can take these signals into account. All these considerations make real-time recommendations with in-session signals as much of a large-scale engineering task as it is a machine learning modeling task. Fig. 1 gives a high-level diagram of such a production system.

\begin{figure*}[h]
  \centering
  \includegraphics[width=.7\textwidth]{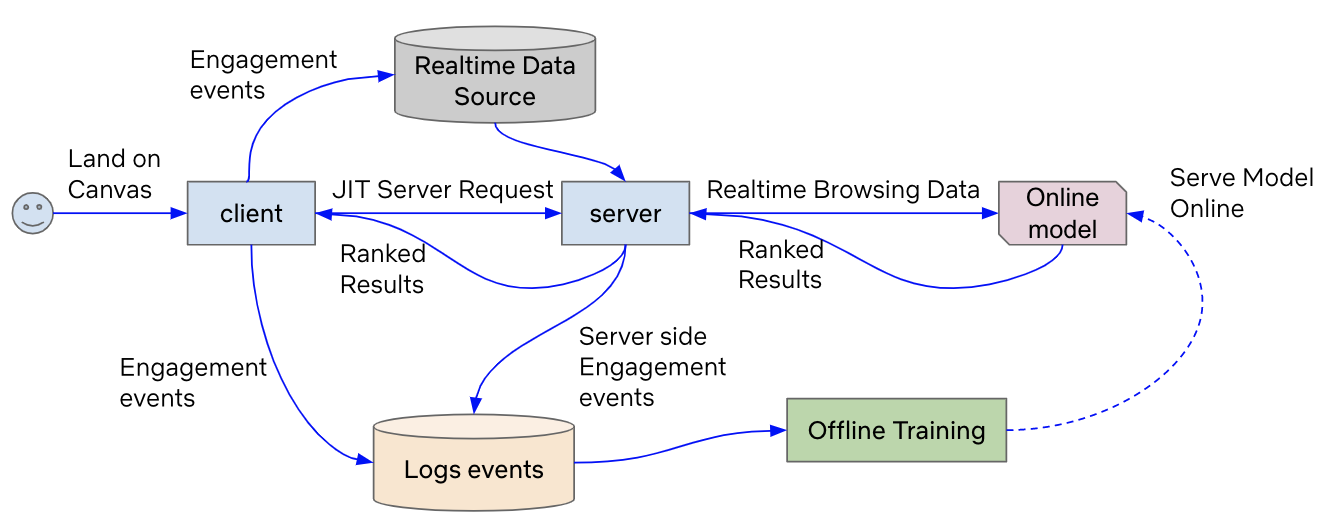}
  \caption{High level overview of an in-session adapted recommendation system for search pre-query.}
\vspace{-6mm}
\end{figure*}

\underline{\textit{Machine Learning Ranker}}: We focus on building a machine learning model that can jointly optimize for the member's long-term and short-term viewing preferences. To do so we took the following approaches:

\begin{enumerate}
\item We develop a personalized ranking model capable of generating relevant recommendations for a member, given her historical engagement with the platform, as well as her \textit{real-time} interaction signals. We consider a few different deep-learning architectures and objectives to jointly learn different types of positive interactions for a given candidate title in a given context. For the in-session adaptation part of the model we use different member interactions observed in the current session to engineer features that can capture information about the member’s in-the-moment intent. For example, we can add signals such as the members’ \textit{genres} and \textit{titles} they interacted positively with during the current session. 

\item We next improve upon the model developed above by taking into account the member’s temporal interaction sequence. Specifically, we can leverage the member’s raw interaction sequence during the current and past few sessions (\textit{cross-session adaptation}). Although the above approach successfully captures the member’s short-term as well as long-term intents, we hypothesize that the raw sequence could enable the model to learn feature representations and signals that the carefully engineered features might have missed. For this stage we experiment with multiple different sequential models, including simple RNN, LSTM, bi-directional LSTM and transformer architecture to leverage a member’s interaction sequence. 

\end{enumerate}

\underline{\textit{Offline Results}}: Our offline analysis shows that an in-session adaptation model significantly outperforms (6\% relative increase in offline ranking metrics) the current production model for the pre-query. Fig. 2 show an example of recommendations generated from the model for pre-query where the member’s pre-query page within the current session updates to capture her most recent interest based on her in-session browsing signals.

\begin{figure*}[h]
  \centering
  \includegraphics[width=.7\textwidth]{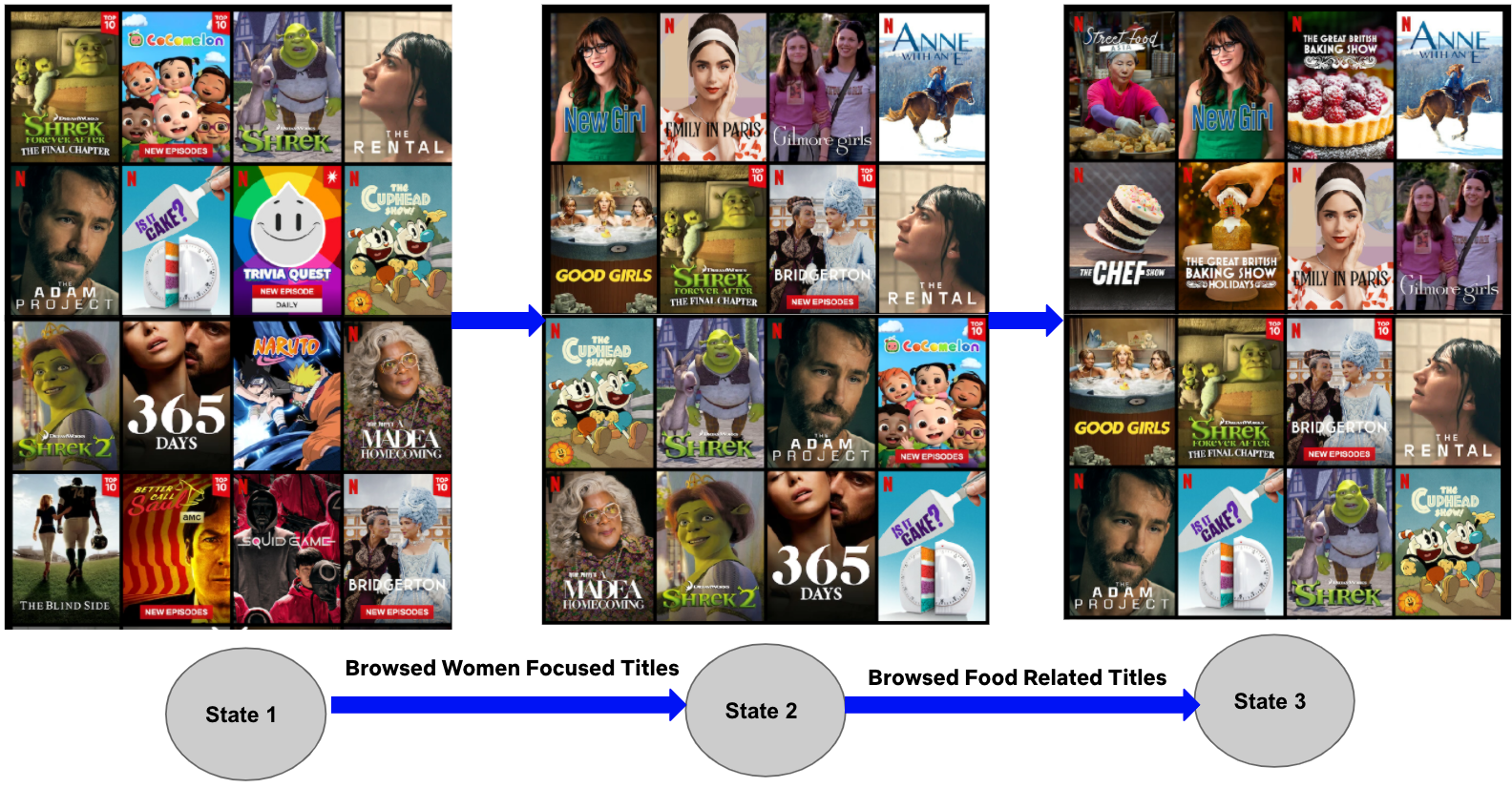}
  \caption{Example recommendations generated from an in-session adapted model for pre-query canvas on Netflix.}
  \vspace{-6mm}
\end{figure*}


\section{Challenges and Future Work }
The member signals leveraged for the approaches developed capture the member’s current interest state and enable the model to identify what the member wants to watch now. However, we still balance it with information about the member’s historical preference, since lack of long-term preference information can result in a limited level of personalization, where a small set of popular items may be recommended to many members. Moreover, we hypothesize that such repeated exposure of a subset of the titles through only a short-term intent prediction model can lead to a concentration effect over time \cite{ferraro2020exploring}. However, a model that jointly optimizes for both the short-term and long-term intent can avoid such myopic recommendations while still catering to the member’s current needs. This is especially important for sessions where there is no prior browsing before landing on search. Additionally, such an adaptive model can improve user cold starting. We faced several challenges in developing an end-to-end in-session adapted recommender systems:

\begin{enumerate}
 \item To generate fresh recommendations that use interaction signals from the current session we cannot precompute and cache recommendations on the server or on the client devices. This may significantly increase the number of calls to the recommender thus increasing timeouts and an increase in infrastructural cost, putting severe constraints on runtime model complexity. Moreover, the increased number of server calls, unreliable or slow networks can degrade the member experience. 

 \item In-session adaptation can make the recommendations too dynamic in some scenarios. Members may not always prefer such fast-changing rankings. For developers it becomes harder to reproduce and debug issues to understand why certain recommendations are showing up. 

 \item Browsing signals from the ongoing sessions can yield extremely sparse features, which needs to be taken into account when developing the model. 
\end{enumerate}

\section{Conclusion}
Leveraging member's near real-time browsing engagements can effectively capture their short-term interest, which can generate relevant pre-query recommendations by anticipating member's search intent. 


\section{Acknowledgments}
We are thankful to our collaborators Rein Houthooft, Vickie Zhang, Weidong Zhang and Christoph Kofler as well as internal reviewers Justin Basilico, Aish Fenton, Vito Ostuni and Yves Raimond.

\section{Bio}
\textbf{\textit{Moumita Bhattacharya}} is a senior research scientist at Netflix where she develops at-scale machine learning models for Search and Recommendation Systems. Prior to Netflix, she was a Senior Applied Scientist at Etsy. She is also an adjunct faculty in the Data Science Institute of University of Delaware. 

\textbf{\textit{Sudarshan Lamkhede}} is the Engineering Manager of the Machine Learning - Search and Recommendations in Netflix Research. Prior to Netflix, he led research engineering for Web Search algorithms at Yahoo! Research. He co-organizes the San Francisco Bay Area Machine Learning symposium (BayLearn).

\bibliographystyle{ACM-Reference-Format}
\bibliography{main}


\end{document}